\documentclass[preprint]{aastex}

\usepackage{epsfig}
\usepackage{amssymb}
\usepackage{amsmath}
\usepackage{rotating}
\usepackage{color}
\usepackage{graphicx}

\slugcomment{accepted by Ap J.}

\shorttitle{On the Origin of Wind Line Variability in O Stars}
\shortauthors{Massa \& Prinja}

\begin{document}
\newcommand{\teff}{{$T_{\rm eff}$}}
\newcommand{\logg}{{$\log g$}}
\newcommand{\iue}{{\it IUE}}
\newcommand{\fuse}{{\it FUSE}}
\newcommand{\ha}{H$\alpha$}
\newcommand{\mdot}{$\dot{M}$}
\newcommand{\mdotq}{$\dot{M} q$}
\newcommand{\spit}{{\it Spitzer}}
\newcommand{\hst}{{\it HST}}
\newcommand{\civ}{C~{\sc iv}}
\newcommand{\ciii}{C~{\sc iii}}
\newcommand{\nv}{N~{\sc v}}
\newcommand{\niii}{N~{\sc iii}}
\newcommand{\niv}{N~{\sc iv}}
\newcommand{\nivline}{N~{\sc iv}~$\lambda$1718}
\newcommand{\ovi}{O~{\sc vi}}
\newcommand{\siiv}{Si~{\sc iv}}
\newcommand{\siivline}{Si~{\sc iv}~$\lambda\lambda$1400}
\newcommand{\sv}{S~{\sc v}}
\newcommand{\svi}{S~{\sc vi}}
\newcommand{\oiv}{O~{\sc iv}}
\newcommand{\ov}{O~{\sc v}}
\newcommand{\heii}{He~{\sc ii}}
\newcommand{\kms}{km s$^{-1}$}

\title{On the Origin of Wind Line Variability in O Stars}


\author{D. Massa}
\affil{Space Science Institute, 4750 Walnut Street, Suite 205, Boulder, CO 
80301}
\email{dmassa@spacescience.org}

\and

\author{R. K. Prinja}
\affil{Department of Physics and Astronomy, University College 
London, Gower Street, London WC1E 6BT}

\begin{abstract}
We analyze 10 UV time series for 5 stars which fulfill specific sampling 
and spectral criteria to constrain the origin of large-scale wind structure 
in O stars.  We argue that excited state lines must arise close to the 
stellar surface and are an excellent diagnostic complement to 
resonance lines which, due to radiative transfer effects, rarely show 
variability at low velocity.  Consequently, we splice dynamic spectra of 
the excited state line, \nivline, at low velocity to those of \siivline\ at 
high velocity in order to examine the temporal evolution of wind line 
features. These spliced time series reveal that nearly all of the features 
observed in the time series originate at or very near the stellar surface.  
Further, we positively identify the observational signature of equatorial 
co-rotating interaction regions in two of the five stars and possibly two 
others.  In addition, we see no evidence for features originating further 
out in the wind.  We use our results to consolidate the fact that the 
features seen in dynamic spectra must be huge, in order to remain in the 
line of sight for days, persisting to very large velocity and that the 
photospheric footprint of the features must also be quite large, $\sim$ 15 
-- 20\% of the stellar diameter.  
\end{abstract}

\keywords{stars: winds, outflows, stars: massive, stars: mass-loss, 
stars: activity}

\section{Introduction}
Revealing the systematic spectral variability of OB star wind lines was a 
major contribution of the {\it International Ultraviolet Explorer} (\iue) 
(e.g., Massa et al.\ 1995, Kaper et al.\ 1996, k96 hereafter).  These 
observations demonstrated that the source of the variability was large, 
coherent structures in the winds (e.g., Fullerton et al.\ 1997, Kaper et 
al.\ (1999, k99 hereafter), de Jong et al.\ 2001, Prinja et al.\ 2002).  
Since then, similar evidence for large scale structure has been detected in 
the winds of LMC O stars (Massa et al.\ 2000), the central stars of 
planetary nebulae (Prinja, Massa \& Cantiello 2012), and may be a 
fundamental property of radiatively driven wind flows.  

While the absorption features observed in time series propagate 
monotonically to higher velocity, they move more slowly than the 
velocity law of the wind (determined from the shapes of saturated wind 
lines) would predict.  In fact, one often observes different features 
propagating with different accelerations at the same time.  To explain this, 
Mullan (1984) proposed that the absorption features were the signatures 
of co-rotating interaction regions, CIRs, in the winds.  These are similar 
to structures seen in the Solar wind, where different sectors of the stellar 
surface give rise to winds which accelerate differently.  Rotation causes 
these different sectorial flows to interact, and an interface with a spiral 
pattern can result and a velocity plateau can develop along the interface.  
Then, as the CIR moves across our line of sight to the stellar disk, it 
can give the appearance of an absorption feature moving slowly outward 
in the wind.  This explanation has considerable appeal, since it explains 
the slow acceleration of the absorption features, and the large spatially 
coherent structures can explain the persistence of the absorption 
features along the line of sight, which can persist for several days.  

Cranmer \& Owocki (1996) and Lobel \& Blomme (2008) analyzed the time 
dependent signature of CIRs originating in the equatorial plane of a star 
when viewed equator-on.  They showed that a specific wavelength dependent 
form of variability resulted, and they termed its shape ``bananas''.  This 
signature has been identified in the \siivline\ wind lines of the B 
supergiants HD~64760 (Fullerton et al.\ 1997) and HD~150168 (Prinja et al.\ 
2002), and the O star $\xi$~Per (de Jong et al.\ 2001).  In these cases, a 
photospheric origin for the variability seems secure.  In fact, 
Ramiaramanantsoa et al.\ (2014) detected optical photometric variability in 
$\xi$~Per with a time scale similar to the DACs measured by de Jong et al., 
strengthening the case for a photospheric connection.  In other cases, it 
is not clear whether the absorption features which appear at low velocity 
and propagate to high velocity actually originate near the stellar surface, 
or if they are the result of a density enhancement caused by low velocity 
material behind a shock further out in the wind, which then simply appears 
at low velocity.  

The presence of structure can significantly affect the mass loss rate 
inferred from different spectral diagnostics and has been the focus of some 
recent observational (e.g., Fullerton et al.\ 2006, Prinja \& Massa 2010, 
Prinja \& Massa 2013) and theoretical (e.g., Sundqvist et al.\ 2010, 
S\v{u}rlan et al.\ 2012) studies.  The result has been a mass loss rate 
discrepancy, which refers to the large, up to a factor of 10, differences 
between the theoretical expectations (e.g., Muijres et al.\ 2012) and 
observationally determined mass loss rates derived from homogeneous wind 
models.  Recent theoretical models have resolved much of this discrepancy 
by introducing optically thick structures in the wind.  However, current 
models are based on winds containing randomly distributed, small scale 
blobs, and the results could be very different for large structures, such 
as CIRs.  The reason is that {\em geometry matters} for wind line formation 
in a wind containing large scale optically thick structures.   

Determining whether the variability seen in the time series is due to 
material originating near or at the surface of the star and then 
propagating outward, or low speed material at large distances from the 
star which simply appears at low velocity is the goal of this paper.  
Should the origin of all of the features be tied to the stellar surface, 
this would have implications about the structure of the winds (how does 
the component of the wind which shapes saturated wind lines co-exist with 
large scale structure), the structure of the photosphere (what causes 
the photospheric irregularities that give rise to the different 
wind flows), X-ray production (do X-rays originate mainly in small, 
spatially incoherent shocks or along the interfaces of the CIRs) and the 
theoretical derivations and simulations based on a smooth wind.  

In the following sections we: 1) provide a brief review of wind line 
formation and variability, emphasizing how excited state lines can be 
used to trace structures to the lower wind and to make a photospheric 
connection; 2) examine the available data sets for O stars with a well 
developed excited state line and arrive at our sample of time series; 3) 
demonstrate one new instance of a CIR in a time series for 68~Cyg, and  
possible CIR signatures in time series for $\zeta$ Pup and $\lambda$~Cep, 
and; 4) discuss the implications of our results.

\section{Wind Line Formation}

When discussing wind line variability, we confine ourselves to the UV lines 
and concentrate on the wind absorption.  Variable emission in UV wind lines 
is rarely observed in OB stars, primarily because the emission at a given 
velocity is composed of light scattered throughout the wind, averaging over 
any inhomogeneities that may be present.  In contrast, the absorption 
arises in the cylinder of wind material that lies between the stellar disk 
and the observer.  The observed absorption is a function of wavelength or, 
equivalently, velocity.  

Generally, there is no way to directly determine the radial location of 
an absorption feature at a specific velocity.  For example, a low velocity 
feature could arise from material near the stellar surface, accelerating 
outward, or from slowly moving wind material far out in the wind, which is 
re-accelerating after being shocked.  This ambiguity is always the case for 
resonance lines, but not for excited state lines.  To see why this is so, 
we now provide a brief description of how the two different lines are 
formed.

A resonance wind line is effectively a pure scattering line.  Anywhere 
an ion exists in the ground state, the resonance line will scatter 
incoming radiation.  For a typical wind acceleration law, a resonance line 
will first become optically thick at low velocity, where the wind density 
is highest.  As a result, wind lines that are strong enough to show 
variability at intermediate and high velocity are usually optically very 
thick at low velocity.  Consequently, large changes in column density 
translate into small changes in the absorption.  Two other factors also 
conspire to make the variability of resonance lines difficult to observe at 
low velocity.  The first is that the photospheric resonance line is 
typically very strong, so any line of sight absorption is seen against a 
very weak continuum.  The second is that the scattered radiation 
originating from all over the wind peaks at low velocity.  This means that 
any weak variation in the small residual flux at low velocity is masked by 
strong emission.   

An excited state wind line arises from an allowed transition whose lower 
level is the upper level of a resonance transition (typically below 
900\AA).  They frequently appear as wind lines in O stars with strong 
mass fluxes.  One of the most commonly observed excited state lines is 
\nivline, whose lower level is the upper level of the \niv$\lambda 765$ 
resonance line.  In most cases, a very simplified level population model 
can be used to accurately describe the population of the lower state of an 
excited state wind lines (e.g., Olson, 1981).   H$\alpha$ (and to a lesser 
degree He~{\sc ii}~$\lambda$1640) is a notable exception.  Because H~{\sc i} 
is a trace ion in O star winds and lies on the only path to the ground 
state, its level populations are dominated by recombinations.  An excited 
state line like \nivline\ has three important properties that make it a 
valuable diagnostic for variability studies.  First, because a strong EUV 
radiation field is required to populate its lower level, it can only exist 
close to the star (this gives excited state lines their distinctive 
profiles).  As a result, {\em a feature which appears at low velocity in 
\nivline\ must originate close to the stellar surface}, i.e., it cannot be 
a low velocity, post-shock region far out in the wind.  Second, the 
photospheric \nivline\ line tends to be considerably weaker than a 
resonance line, so a variable absorption is viewed against a stronger 
continuum than for a resonance line.  Third, because excited state lines 
become optically thin far from the star, the scattered light contribution 
at low velocity is much weaker.  These properties make variability much 
easier to detect at low velocity in an excited state line.  However, 
because excited state lines weaken quickly at large distances from the 
star, high velocity variability is difficult to detect.  

We see, therefore, how resonance and excited state lines complement one 
another.  If a feature (indicated by excess or reduced absorption) appears 
at low velocity in an excited state line and then joins a high velocity 
feature in a resonance line (whose low velocity portion may be saturated) 
at higher velocity, this provides evidence that the excess or deficiency of 
opacity which caused the feature close to the stellar surface (where the 
radiation field is intense) propagated outward, into the wind.

\section{The Data}
\begin{figure}
\begin{center}
\vspace{-1.0in}\includegraphics*[width=6.5in]{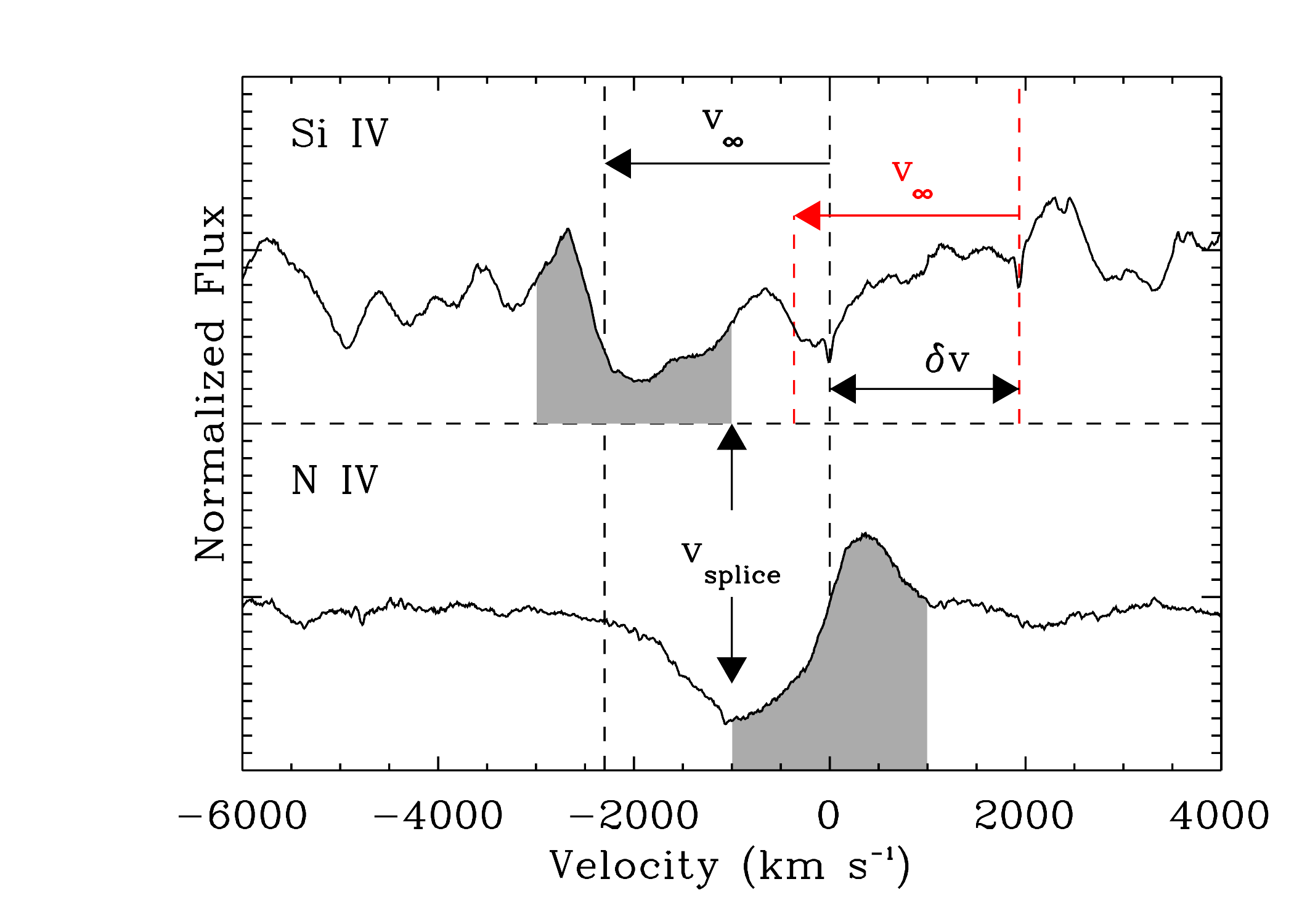}
\end{center}
\caption{Cartoon depicting the velocity ranges used to splice together the 
combined \siiv\ and \niv\ dynamic spectrum.  The figure shows the mean 
spectrum from one of the $\zeta$~Pup time series for the \siiv\ region 
(top) as a function of velocity relative to the blue component of the 
doublet and the \niv\ region (bottom) as a function of velocity relative 
to its rest wavelength.  Arrows indicate the ranges of the doublet profile 
affected by the wind, and the doublet separation.  The vertical arrow 
indicates the velocity at which the two profiles are ``stitched'' 
together.\label{fig:cartoon}}
\end{figure}

The \iue\ data base is the only one available containing the UV time series 
needed for analysis.  To avoid radiative transfer effects, one would prefer 
to compare a resonance singlet and an excited state singlet.  While 
\nivline\ is an ideal example of the latter, O star spectra do not have a 
suitable resonance singlet in the \iue\ wavelength band.  Consequently, we 
were forced to use a doublet.  In this case, \siivline\ is the best 
candidate, since its doublet separation is large, $\delta v = 1936~$\kms, 
and it is often strong but unsaturated in stars with well developed 
\nivline.  We concentrate on the blue component of the doublet because 
it is unaffected by the red component over the velocity range $-v_\infty 
\leq v \leq -v_\infty +\delta v$ (Olson 1982).  Thus, much of the high 
velocity portion of the blue component behaves as if it were a singlet.  
For example, the blue component of the \siiv\ doublet in a star with 
$v_\infty = 2700$~\kms\ is unaffected by the red component over the 
velocity range $-2700 \leq v \leq -764$~\kms.  Figure~\ref{fig:cartoon} 
shows these intervals for $\zeta$~Pup.  

To arrive at our sample, we began with all the SWP high resolution 
observations of stars in \iue\ object classes 12--15, which stand for: main 
sequence O; supergiant O; Oe, and; Of.  This consisted of 3401 spectra of 
325 stars.  The sample was then narrowed to stars with 10 or more spectra 
(54 stars) with sequential observations separated by less than 1 day (27 
stars).  We then imposed the following two constraints.  First, the spectra 
must have distinctly asymmetric \nivline\ profiles, so that wind 
variability can be observed.  Second, the series must cover 4 or more days 
to span enough time to determine whether features observed at low velocity 
in \niv\ appear later at high velocity in \siiv.  The net result of this 
search was 12 potentially suitable time series of 6 stars.  

Table~\ref{tab:sample} gives some basic properties of the stellar sample. 
The columns give: 1 -- The star name, 2 -- spectral types from 
Ma{\'i}z-Apell{\'a}niz et al.\ (2003), 3 -- wind terminal velocities from 
Prinja et al.\ (1998) for HD~93843 and the compilation of Fullerton et 
al.\ (2006) for the rest, 4 -- projected rotational velocities from Penny 
(1996), 5 -- stellar radii (again from Fullerton et al.\ or Prinja et 
al.), and 6 --  the number of time series inspected.

Table~\ref{tab:series} summarizes the properties of the time series we will 
examine.  The columns list: 1 -- the star name, 2 -- the duration of the 
series in days, 3 -- the range of \iue\ high resolution image numbers which 
encompass the series, 4 -- the number of spectra in the series, 5 -- the 
dates spanned by the series, and 6 -- comments on the individual series.

Unfortunately, a few of the time series are not very useful for our 
purposes.  For example, the series for HD~93843 is a long, but very 
sparsely sampled series (see Prinja et al.\ 1998), with a mean sampling 
interval of 10.2~hr.  Such sparse sampling makes it difficult to identify 
features at low velocity, where they are accelerating quickly.  
To quantify this, consider the time it takes a parcel of wind to accelerate 
from its initial velocity, $v_1$, to a specific velocity, $v_2$. We adopt a 
standard $\beta$ law of the wind acceleration, which has the form, 
$w = (1 -a/x)^\beta$, where $w = v/v_\infty$, $x = r/R_\star$, and $a = 1 
-w(x=1)$.  Adopting $\beta = 1$ (a common value for O stars), we find that 
the time it takes a parcel of wind material to accelerate from $w_1$ to 
$w_2$ is: 
\begin{equation}
t_2 -t_1 = 193.3\; \frac{aR_\star}{v_\infty} 
\left[ \frac{w_2-w_1}{(1-w_1)(1-w_2)} +\ln \frac{w_2(1-w_1)}{w_1(1-w_2)}
\right] \; {\rm hrs}\label{eq:flow}
\end{equation}
Equation~(\ref{eq:flow}) shows that it takes wind material $\sim 4.9$ hours 
to accelerate from $w = 0.02$ to $w = 0.5$ (where features tend to be lost 
in \niv).  While absorption features typically accelerate more slowly than 
this, the sampling is so sparse that it can easily miss the entire time 
evolution of low velocity \niv\ features.  As a result, we eliminate the 
HD~93843 series from the following analysis.  For similar reasons, we also 
drop the third series for $\lambda$~Cep.

\section{Analysis}
\begin{figure}
\vspace{-1.0in}\includegraphics*[width=5.5in, angle=90]{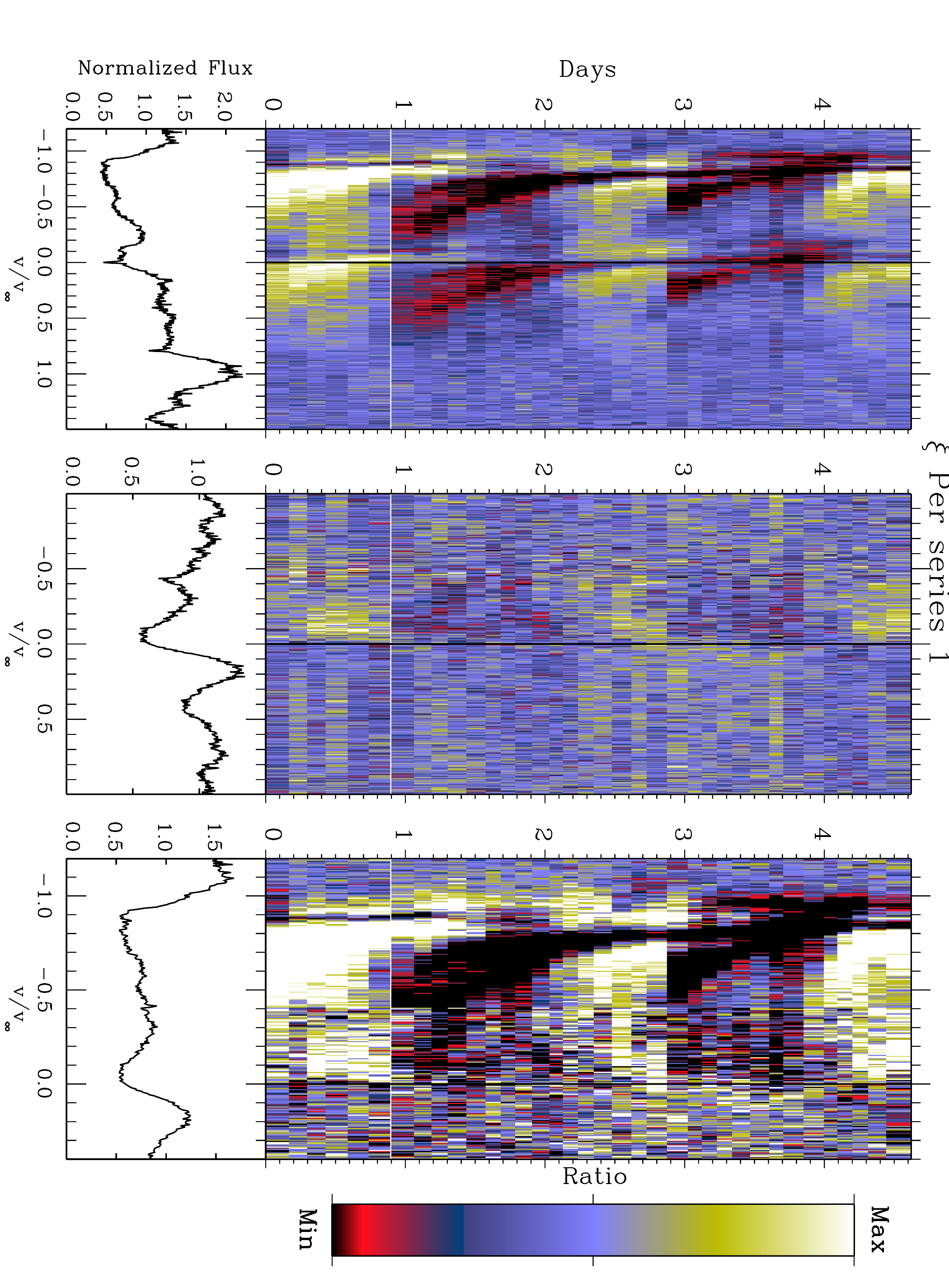}
\caption{Dynamic spectra for \siiv\ (left), \niv\ (center) and the spliced 
spectrum (right).  Each is plotted against $w = v/v_\infty$, where the 
$v_\infty$ are from Table~\ref{tab:sample}.  The dynamic spectrum on the 
left is for \siivline\ relative to the rest velocity of the red component, 
1402.77\AA, and with limits of 1/1.4 and 1.4.  The center dynamic spectrum 
is for \nivline\ with limits of 1/1.2 to 1.2.  The right is for the spliced 
line and also has limits of 1/1.2 to 1.2.  Normalized versions of the mean 
spectra used to normalize the images are shown below each one.
\label{fig:xiper1}}
\end{figure}

\begin{figure}
\vspace{-1.0in}\includegraphics*[width=5.5in, angle=90]{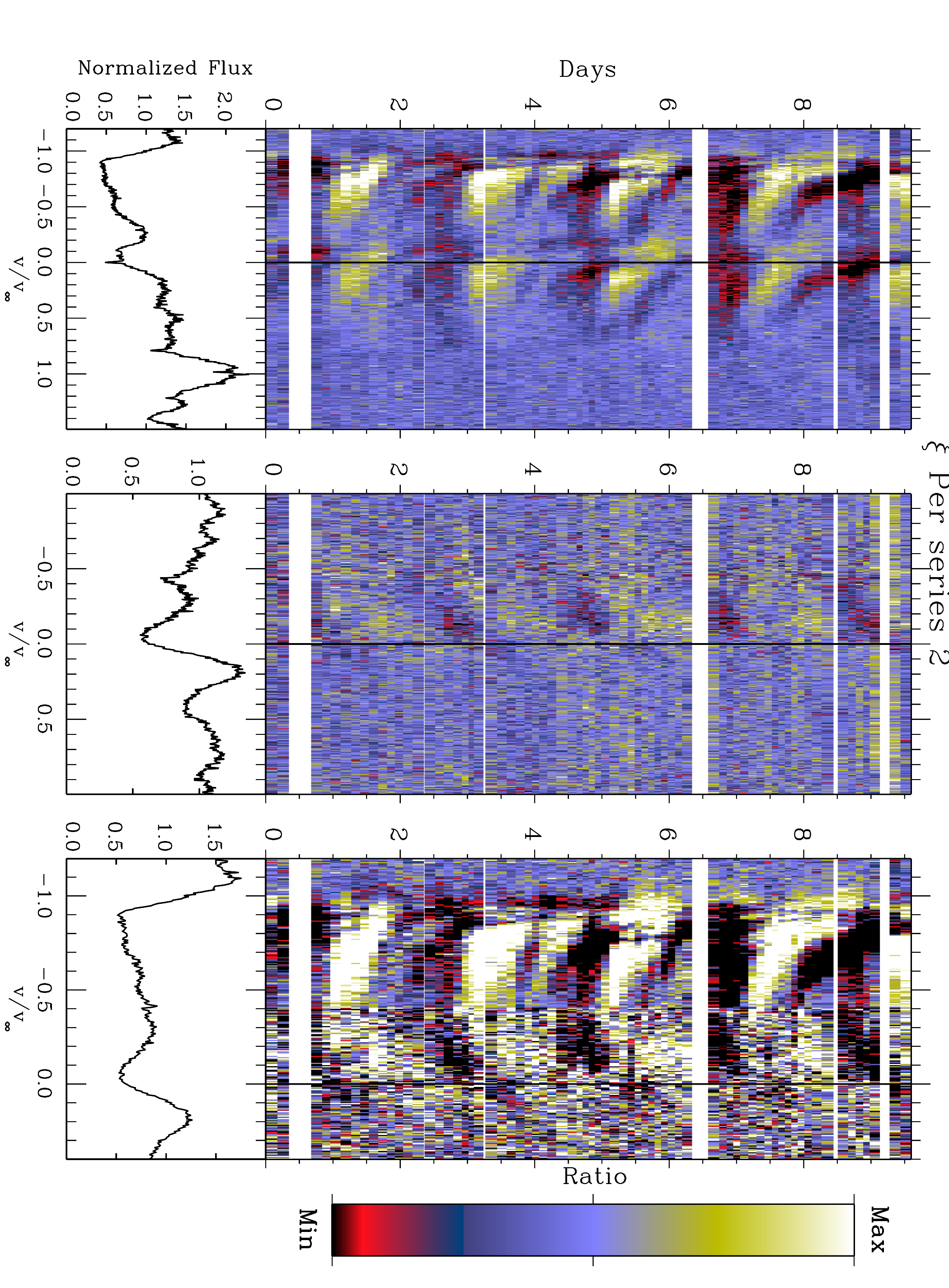}
\caption{Same at Figure~\ref{fig:xiper1} for the second $\xi$~Per time 
series.\label{fig:xiper2}}
\end{figure}

\begin{figure}
\vspace{-1.0in}\includegraphics*[width=5.5in, angle=90]{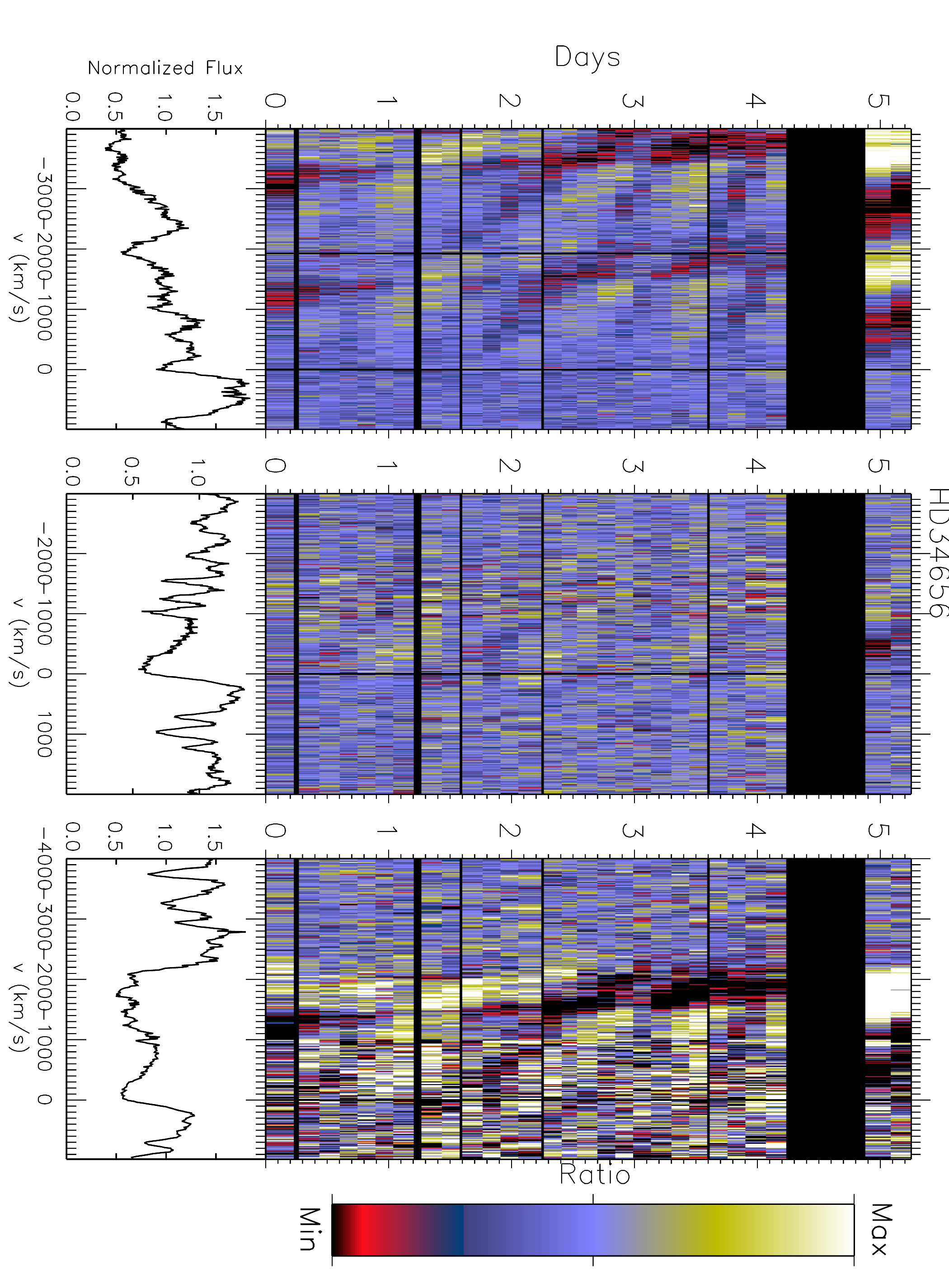}
\caption{Same at Figure~\ref{fig:xiper1} for the HD~34656 time series.
\label{fig:hd34656}}
\end{figure}

\begin{figure}
\vspace{-1.0in}\includegraphics*[width=5.5in, angle=90]{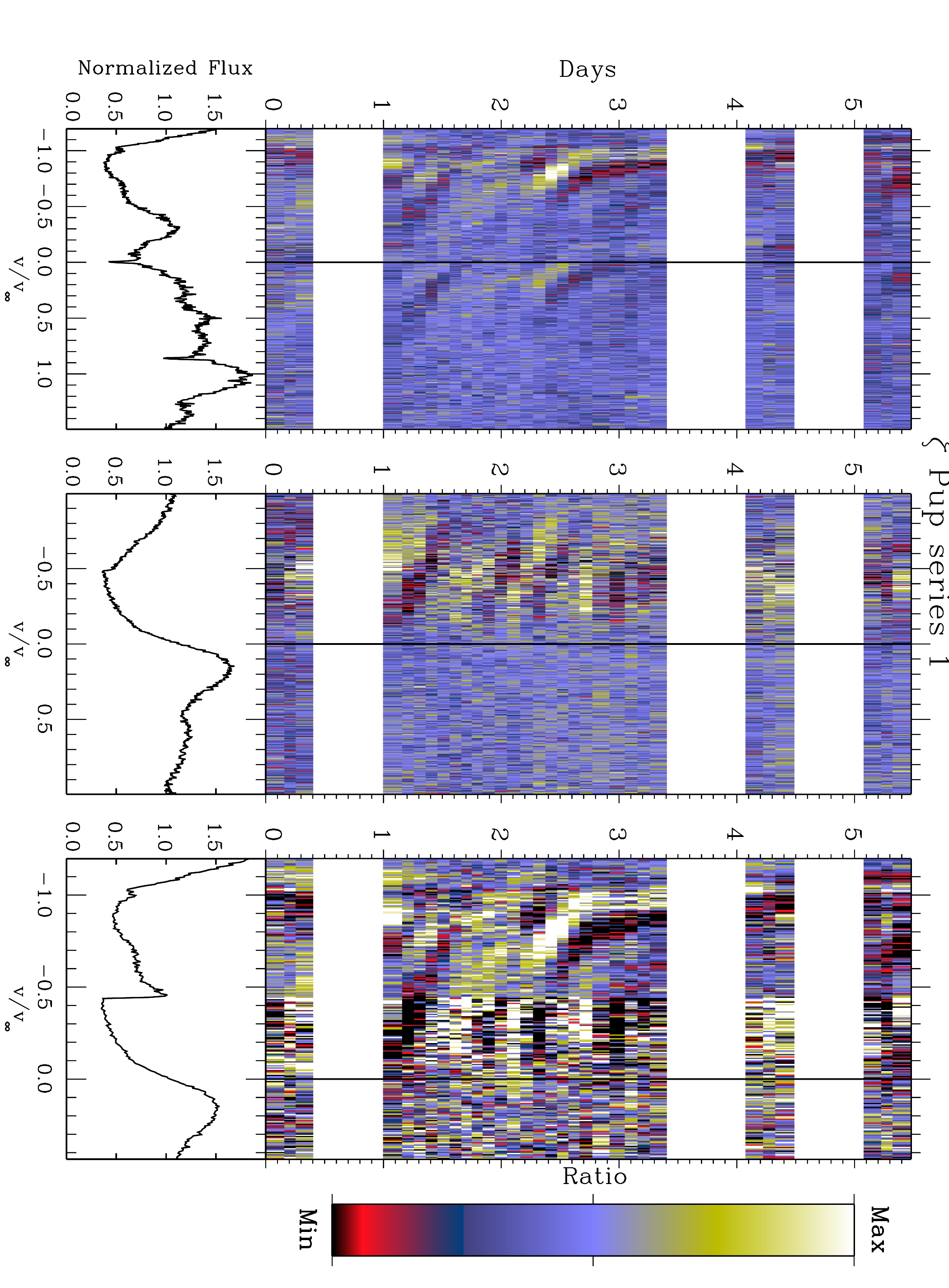}
\caption{Same at Figure~\ref{fig:xiper1} for the first $\zeta$~Pup time 
series.\label{fig:zetpup1}}
\end{figure}

\begin{figure}
\vspace{-1.0in}\includegraphics*[width=5.5in, angle=90]{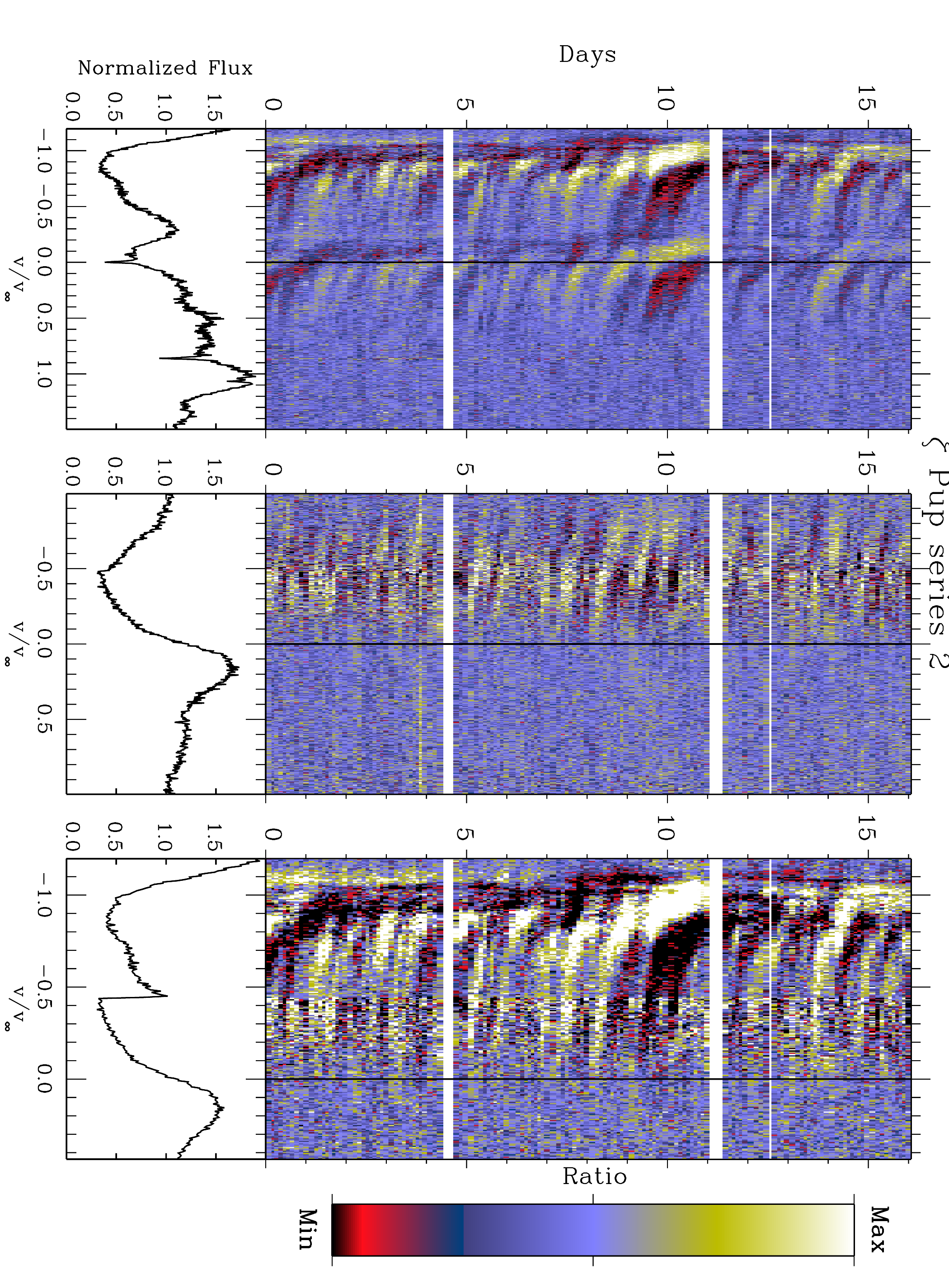}
\caption{Same at Figure~\ref{fig:xiper1} for the first $\zeta$~Pup time 
series.\label{fig:zetpup2}}
\end{figure}

\begin{figure}
\vspace{-1.0in}\includegraphics*[width=5.5in, angle=90]{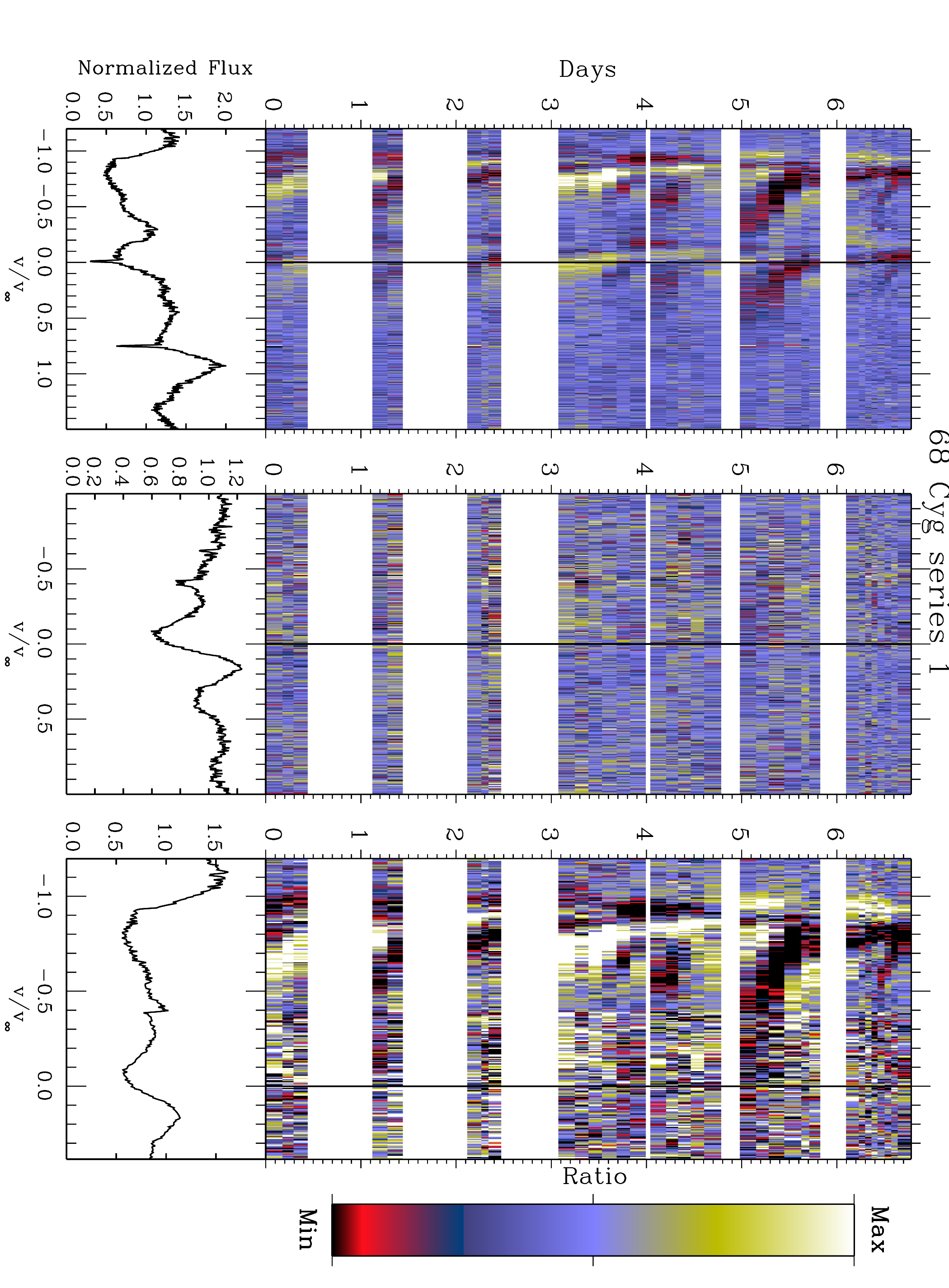}
\caption{Same at Figure~\ref{fig:xiper1} for the first 68~Cyg time 
series.\label{fig:68cyg1}}
\end{figure}

\begin{figure}
\vspace{-1.0in}\includegraphics*[width=5.5in, angle=90]{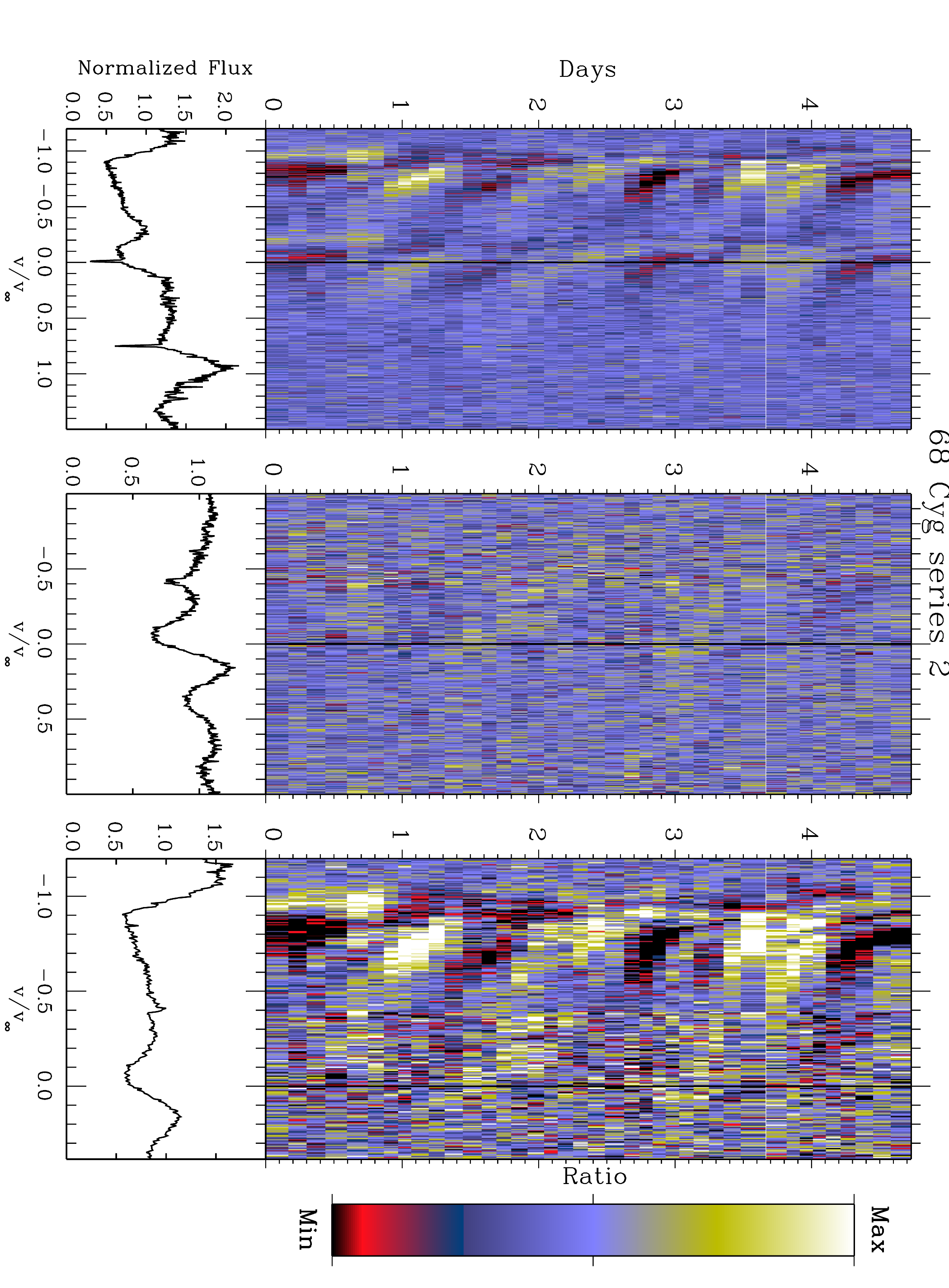}
\caption{Same at Figure~\ref{fig:xiper1} for the second 68~Cyg time 
series.\label{fig:68cyg2}}
\end{figure}

\begin{figure}
\vspace{-1.0in}\includegraphics*[width=5.5in, angle=90]{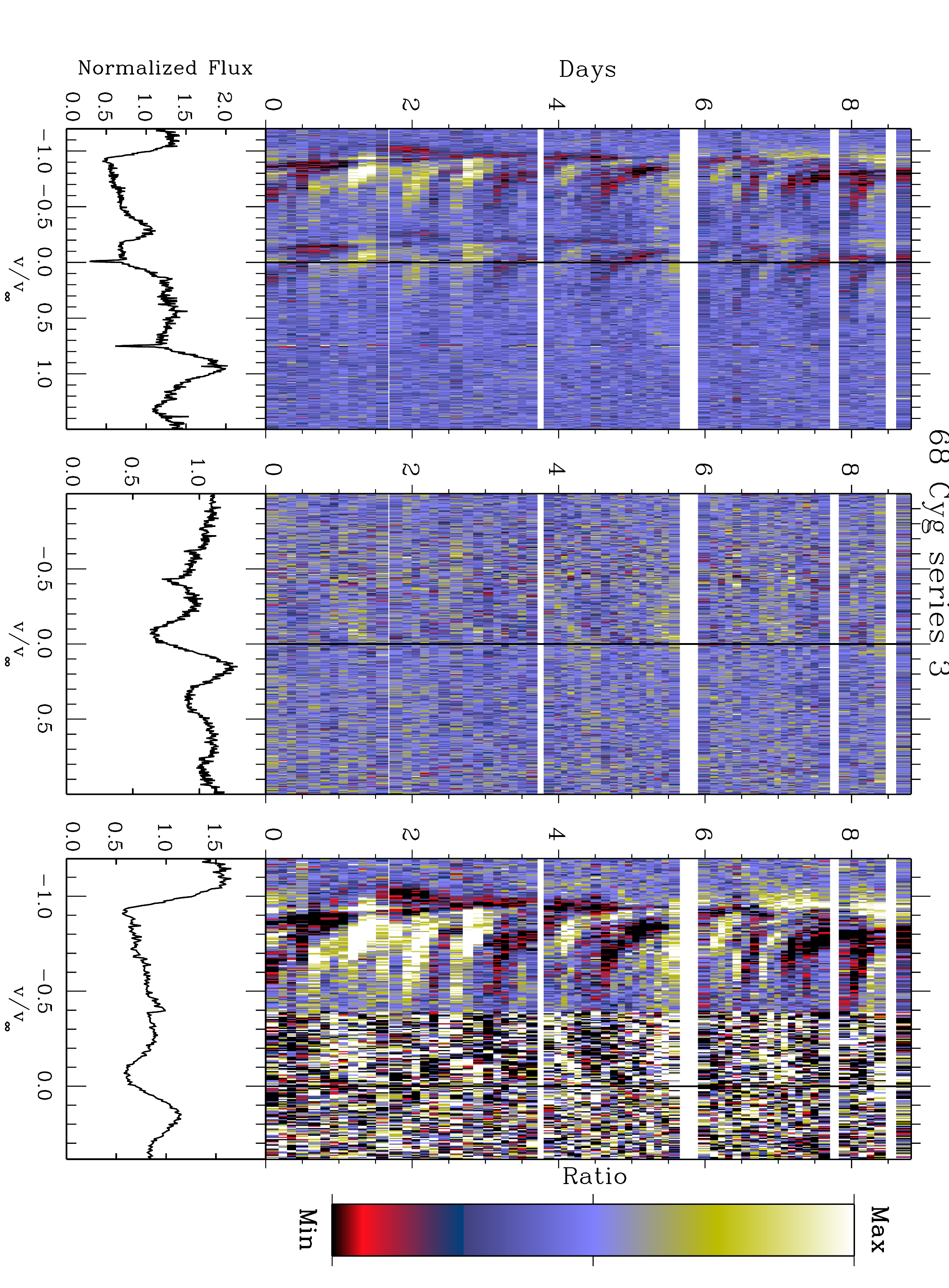}
\caption{Same at Figure~\ref{fig:xiper1} for the first 68~Cyg time 
series.\label{fig:68cyg3}}
\end{figure}

\begin{figure}
\vspace{-1.0in}\includegraphics*[width=5.5in, angle=90]{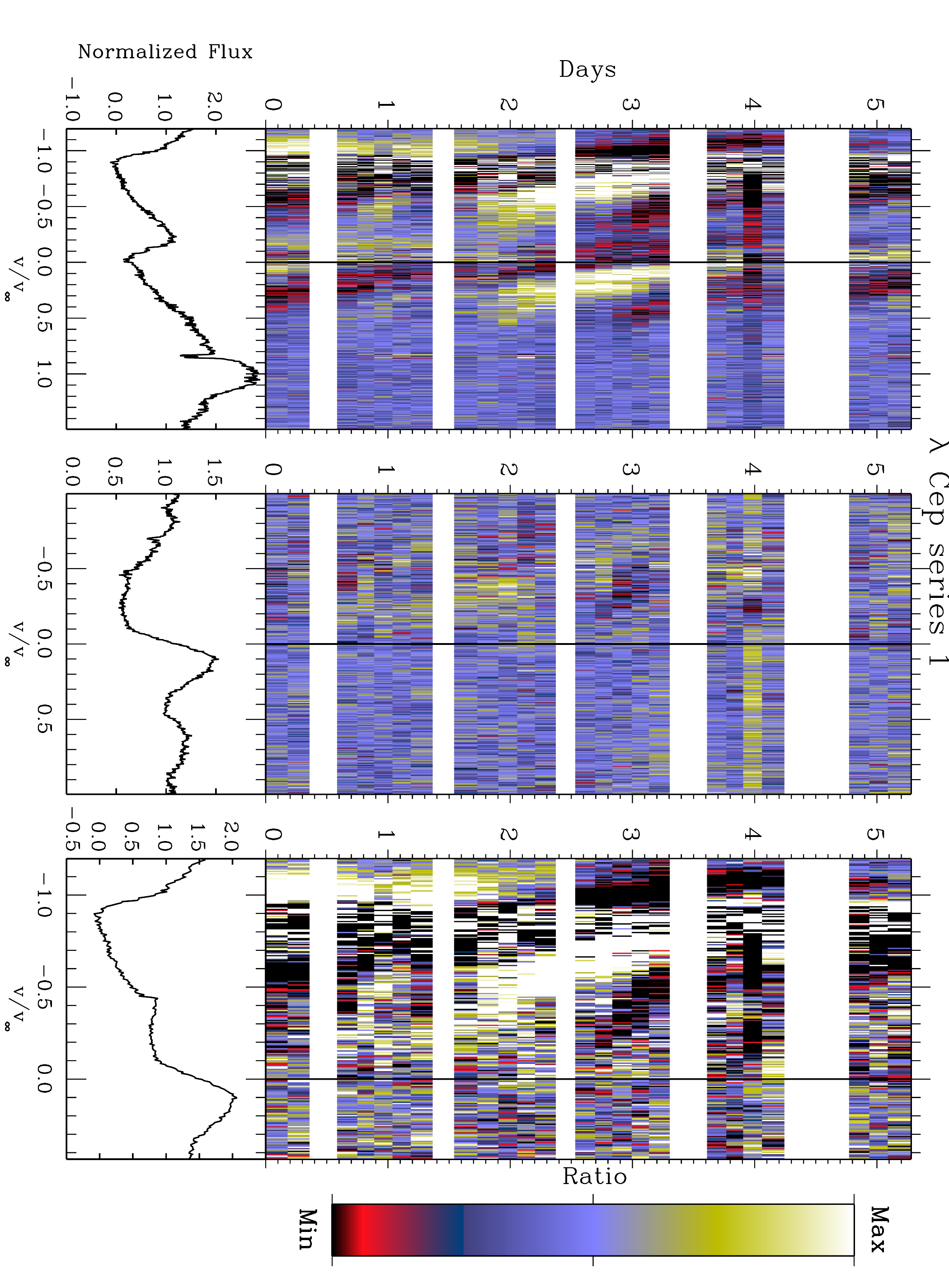}
\caption{Same at Figure~\ref{fig:xiper1} for the first $\lambda$~Cep time 
series.\label{fig:lamcep1}}
\end{figure}

\begin{figure}
\vspace{-1.0in}\includegraphics*[width=5.5in, angle=90]{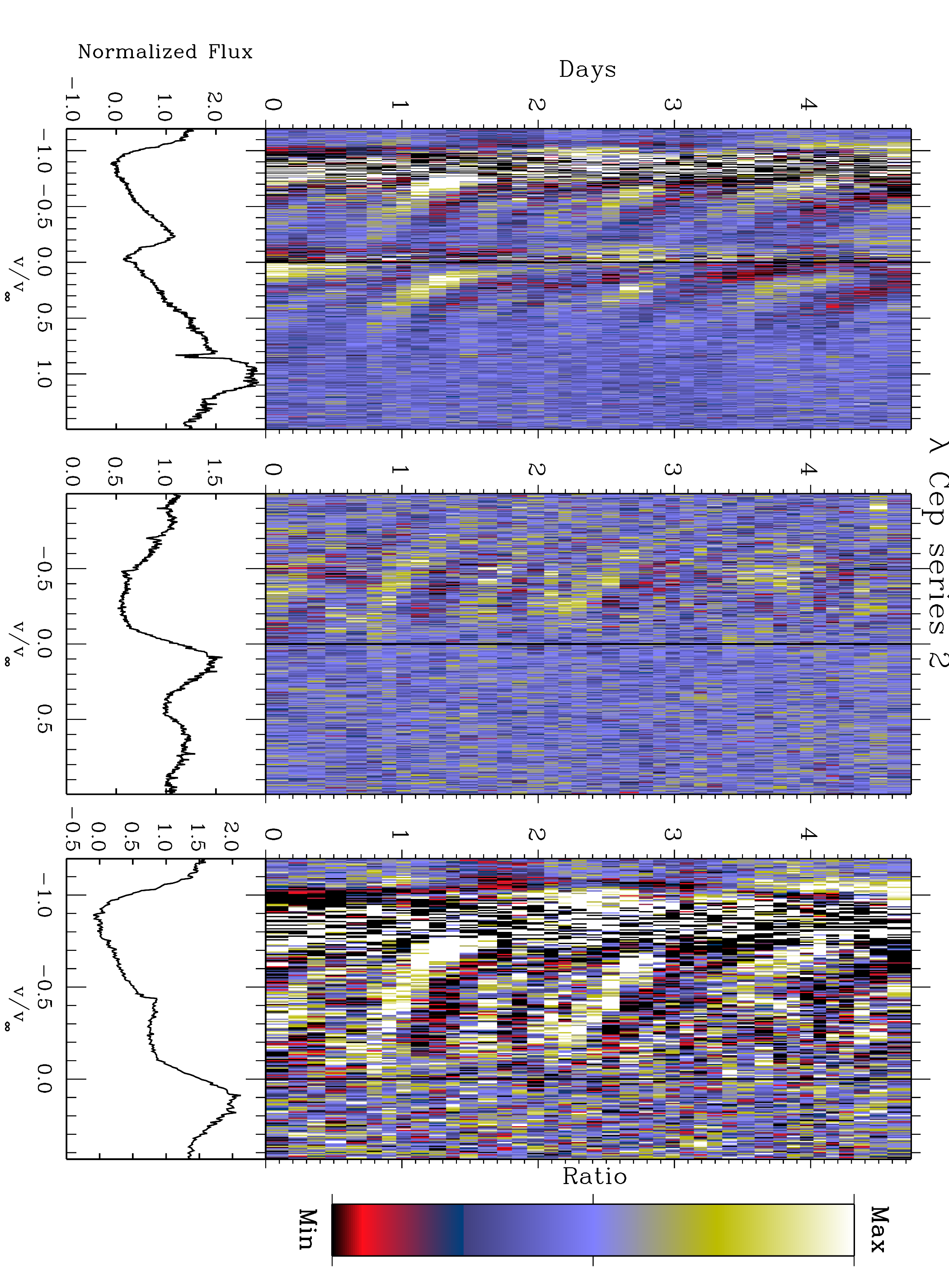}
\caption{Same at Figure~\ref{fig:xiper1} for the second $\lambda$~Cep time 
series.\label{fig:lamcep2}}
\end{figure}

Our analysis is relatively simple, and based primarily on visual 
inspection of dynamic spectra.  Dynamic spectra are images of time ordered 
spectra normalized by the mean spectrum of the series, i.e., $n_\lambda = 
f_\lambda/\langle f_\lambda \rangle$.  Figures~\ref{fig:xiper1} -- 
\ref{fig:lamcep2} show dynamic spectra for the program stars.  Each figure 
contains three dynamic spectra with time increasing upward (labeled by the 
number of days since the beginning of the series) plotted against normalized 
velocity, $w = v/v_\infty$, where the $v_\infty$ values are taken from 
Table~\ref{tab:sample}.  Nearest neighbor interpolation was used to place 
the spectra on a linear time grid, and gaps appear whenever more than 5 
hours elapsed between exposures. The vertical lines represent the rest 
wavelengths of the lines, the bar at the side of each set shows the color 
scale used, and the mean spectrum for the series is shown at the bottom of 
each panel.  The first two panels of each figure are for the \siivline\ 
resonance doublet and the \nivline\ excited state line and the minimum and 
maximum scaling for these are 1/1.4 to 1.4 and 1/1.2 to 1.2 respectively.  
The third panel shows the low speed portion of the \niv\ dynamic spectrum 
($v \leq 1000$~\kms) spliced onto the high speed portion ($v > 1000$~\kms) 
of the \siiv\ one (Figure~\ref{fig:cartoon} shows the regions spliced 
from each line).  We choose 1000~\kms\ so that the high speed portion of the 
blue component of the resonance doublet will be radiatively decoupled from 
the red component.  Because the amplitude of the \niv\ variability near 
1000~\kms\ is typically smaller than the that of \siiv, we rescale the 
\niv\ portion of the profile to make the amplitudes match.  This was done 
as follows:
\begin{equation}
n_\lambda^\prime = (n_\lambda -1) \times Const +1 
\end{equation}
where a value of $Const = 2.5$ was used for all of the time series except 
for $\zeta$~Pup, which has very strong \niv, so a value of 1.5 was used.  
The minimum and maximum extents of the color range for the spliced profiles 
are 1/1.2 to 1.2.   

Our final sample is biased toward rapidly rotating stars.  This is a 
consequence of the findings by Prinja (1988), which determined that wind 
activity evolves more quickly in rapidly rotating stars, so later time 
series observations concentrated on these objects to maximize the activity 
that could be captured in a fixed amount of \iue\ observing time. 

The remainder of this section summaries the result of each spliced time 
series.

\noindent \underline{{\bf $\xi$ Per}} (Figures~\ref{fig:xiper1} and 
\ref{fig:xiper2}): The first series for this star has been examined 
previously by k96, Kaper et al.\ (1997, k97 hereafter), and k99.  The series 
lasts only 4.4 days.  Nevertheless, during that time 2 major absorption 
events occur, and both of them appear to have the ``banana'' shape that is 
indicative of an equatorial CIR moving into the line of sight.

The second series was analyzed extensively by de Jong et al.\ (2001).  It 
is roughly twice as long as the first and, as de Jong et al.\ point out, 
contains the signatures of at least 4 CIR passages. 

\noindent \underline{{\bf HD 34656}} (Figure~\ref{fig:hd34656}): 
Our time series for this star was previously analyzed by k96, k97, and k99.  
The dynamic spectrum contains four major absorption features \siiv, and all 
of these connect to low velocity features in \niv.  
While these features lack the signature of an equatorial CIR viewed edge on, 
at least two of them appear to extend to $v\simeq 0$.  Although faint, is 
seems clear that one can trace absorption features from $(w, t) = (0.0, 1.3) 
\rightarrow (-0.7, 2.4)$, and $(0.0, 2.5) \rightarrow (-0.6, 3.1)$. 

\noindent \underline{{\bf $\zeta$ Pup}} (Figures~\ref{fig:zetpup1} and 
\ref{fig:zetpup2}): 
The first series for this star was analyzed by Prinja et al.\ (1992).   It 
is poorly sampled for our purposes, with large gaps in the time coverage.  
Nevertheless, at least 3 features can be seen moving through the \siiv\ 
profile, and all of them appear to connect to low velocity features in 
\niv\ which extend to $v\simeq 0$.  

The second series was described by Howarth et al.\ (1995).  It is long and 
well sampled.  Numerous features are apparent both \siiv\ and \niv, and all 
of them appear to join seamlessly, indicating that they all originate near 
or at the stellar surface.  A few of the features (e.g., the ones near day 
4, day 9 and day 14) may have CIR signatures.  

\noindent \underline{{\bf 68 Cyg}} (Figures~\ref{fig:68cyg1} -- 
\ref{fig:68cyg3}): This star has the weakest \nivline\ of the sample.  The 
first series has been discussed several times in the past (Prinja 1988, 
Prinja \& Howarth, 1988, k96, k97, k99).  This series is poorly sampled for 
our purposes.  Nevertheless, there are good indications that the high and 
low velocity features join.  Specifically, the feature between days 5 and 6 
and the two between days 3 and 5 appear to connect.  The large gaps in 
coverage for the rest of the series makes it impossible to draw conclusions 
about the remainder of the features seen in \siiv.

The second 68 Cyg series was analyzed by Howarth \& Smith (1995), k96, k97 
and k99.  It is a short, but well sampled series and it is quite 
clear that all of the high velocity features seen in \siiv\ join low 
velocity features in \niv.

The third 68 Cyg series has not been presented previously.  It is fairly 
well sampled and considerably longer than the others.  Once again, in spite 
of the weakness of the \niv\ variability, it is apparent that all of the 
high and low velocity features join.  In fact, an equatorial CIR signature 
is apparent in many of them.   

\noindent \underline{{\bf $\lambda$ Cep}} (Figures~\ref{fig:lamcep1} and  
\ref{fig:lamcep2}): 
The first series for this was included in k96, k97, and k99.  It is poorly 
sampled with large gaps in the coverage.  In spite of the poor coverage, it 
is possible to follow some of the features to zero velocity, and those that 
cannot be followed are due to poor coverage.   

The second $\lambda$~Cep series was also included in k96, k97and k99.  It 
is well sampled and presents an excellent example of distinct high and low 
velocity features joining seamlessly.  The feature that straddles day 1 
hints of having a CIR signature.  

\section{Discussion}
We begin by summarizing our main observational results.  We then discuss 
their implications for the structure of O star winds.  Finally, we 
highlight some issues that must be resolved before we can arrive at a 
complete understanding of OB star winds.  

Our main observational results are the following: 
\begin{enumerate}
  \item Every absorption feature observed at low velocity connects to a 
    feature at high velocity.
  \item Conversely, whenever the data allow it, every feature observed at 
    high velocity can be traced to a feature at low velocity.
  \item Some of the spliced \siiv -- \niv\ dynamic spectra reveal 
    ``bananas'', which are the accepted signatures of CIRs tied to the 
    stellar surface. 
  \item We see evidence of an equatorial CIR in 68 Cyg for the first time, 
    and possible equatorial CIR signatures in spliced time series for 
    $\zeta$~Pup and $\lambda$~Cep. 
  \item Since our technique reveals CIR signatures which almost certainly 
    indicate the presence of spiral arms in the winds, we suspect that 
    the other features which extend to $v \simeq 0$ are also caused by 
    CIRs, just not in the equatorial plane, or viewed edge on. 
  \end{enumerate}

These results have the following implications on the structure of OB star 
winds:
\begin{enumerate}
  \item The fact that features observed at low velocity in excited state 
    lines join seamlessly to features at high velocity lines and vice 
    versa, establishes a connection between the wind features and the 
    stellar surface.
  \item For the series analyzed, we see no evidence for features formed at 
    intermediate velocity in the wind, i.e., all features can be traced 
    back to the surface and they move through the profile monotonically 
    with velocity.  This implies the features are not formed by shocked gas 
    far from the star, but by some mechanism very close to or on the 
    stellar surface.
  \item As has been known for a long time, the features seen in dynamic 
    spectra must be huge, in order to remain in the line of sight for days, 
    persisting to very large velocity.
  \item For a feature to be detected at $v \simeq 0$, it must cause a 2 
    -- 4\% (Howarth \& Smith, 1995) decrease in the flux or, equivalently, 
    occult at least 2 -- 4\% of the stellar disk.  This means that the 
    linear size of the region responsible for the excess absorption must be 
    $\sim$ 15 to 20\% of the stellar diameter and, in some cases, 
    considerably larger.  
\end{enumerate}

While the current results represent real progress, several open questions 
remain and we now describe a few of them.  First, the stars in our sample 
which have been monitored at optical wavelengths show only very small 
amplitude photometric variability (Howarth \& Stevens 2014, 
Ramiaramanantsoa et al., 2014).  Because we have shown that the 
photospheric footprints of the variability must be quite large, this 
implies that either the temperature difference between the footprints and 
the surrounding regions is very small or else the wind structures arise 
from small, randomly distributed irregularities which somehow organize 
themselves into large scale wind structures very near the stellar surface. 
Second, it is currently thought that the X-rays originate from small, 
spatially incoherent shocks distributed throughout the wind (e.g., Cohen 
et al.\ 2014).  However, when monitored in X-rays, $\zeta$~Pup (Naz\'{e} 
et al.\ 2013) showed no evidence for variability on short time scales, 
implying that if the X-rays are produced by small scale structures, there 
must be thousands of them present at any given time.  On the other hand, 
both $\zeta$~Pup and $\xi$~Per (Massa et al.\ 2014) exhibit X-ray 
variability on time scales that are more consistent with the those of the 
large structures we observe in the UV wind lines.  Three possible 
explanations are: 1) the CIRs shepherd the clumps (although exactly how 
this occurs is not clear); 2) the CIRs occult some of the clumps as they 
sweep across the line of sight (but this requires them to be optically 
thick to X-rays and carry considerably more mass than currently believed), 
or; 3) the CIRs themselves are the source of the X-rays and, since they 
are large, coherent structures, short time scale variability is not 
expected (such a model has not been examined).  Whatever the case, the 
fact that X-rays vary on a time scale consistent with the spiral structures 
observed in the UV wind lines highlights a short coming of our current 
understanding how OB star winds are structured.
Third, we know that the UV spectral diagnostics used to derive mass loss 
rates in OB stars are strongly affected by optically thick structures 
(Massa et al.\ 2008, Prinja \& Massa 2010).  Further, Sundqvist et al.\ 
(2010) and S\v{u}rlan et al.\ (2013) have demonstrated how small, optically 
thick clumps can affect the diagnostics.  However, for large scale, 
optically thick structures, geometry matters, i.e., the exact location and 
velocity of the structures affects the resulting wind line profiles, but to 
what extent is not at all clear.  While CIRs may have a small effect on the 
mass loss rates (Lobel \& Blomme 2008), they can have a profound effect on 
the wind lines.  This is because they are thought to form in a velocity 
plateau, which can make their optical depths as much as 100 times larger 
than the ambient wind (see, Cranmer \& Owocki 1996).  Further, CIRs can 
affect a large velocity range at any given time (see, Figure 3 in Owocki et 
al.\ 1995).  Unlike the small scale structures, the effects that optically 
thick CIRs may have on wind lines have not been modeled.  Consequently, it 
is not clear how much of an effect they can have on our interpretations of 
observed profiles.  Finally, we note that according to Cranmer \& Owocki 
(1996), the DACs are caused by a non-monotonic velocity law.  This occurs 
when the faster wind collides with the slower wind.  For this mechanism to 
be responsible for what we observe, the two winds must interact very close 
to the stellar surface.  While this is not typical for CIR formation, it is 
not impossible, and sets a strong constraint on the winds from the two 
regions that create the CIRs.  

\acknowledgments
We thank Alex Fullerton for useful discussions and the referee for helpful 
comments on how to clarify and focus the presentation.  DM acknowledges 
support from NASA's Astrophysics Data Analysis Program through Grant 
NNX14AB30G.  The data presented in this paper were obtained from the 
Mikulski Archive for Space Telescopes (MAST). STScI is operated by the 
Association of Universities for Research in Astronomy, Inc., under NASA 
contract NAS5-26555. Support for MAST for non-HST data is provided by the 
NASA Office of Space Science via grant NNX13AC07G and by other grants and 
contracts.

{}

\clearpage


\begin{table}
\caption{Program Stars}\label{tab:sample}
\begin{center}
\begin{tabular}{llcccc}
Name          &  Sp Ty            & $v_\infty$ & $v\sin i$ & 
$R/R_\odot$ & No.\ Series \\ 
&                   & km s$^{-1}$ & km s$^{-1}$ &  & 
  \\ \hline
$\xi$ Per      &  O7.5 III(n)((f)) & 2450 & 204 & 14.0 &   2 \\
HD 34656       &  O7 II(f)         & 2150 &  85 & 24.1 &   1 \\
$\zeta$ Pup    &  O4 I(n)f         & 2250 & 203 & 19.4 &   2 \\
HD 93843       &  O5 III(f) var    & 2650 & 100 & 14.0 &   1 \\
68 Cyg         &  O7.5 III:n((f))  & 2550 & 295 & 15.7 &   3 \\
$\lambda$ Cep  &  O6 I(n)fp        & 2250 & 214 & 21.1 &   3 \\  \hline
\end{tabular} 
\end{center}
\end{table}

\begin{table}
\caption{\iue\ Time  Series}\label{tab:series}
\begin{center}
\begin{tabular}{lccccl}
Name          &  Duration    & SWP range & No.\ of & Dates & Comments \\  
              &     days     &           & spectra   & yyyy/mm/dd -- mm/dd 
& \\  \hline
$\xi$ Per     &  4.4 & 42788 -- 42918 &  36 & 1991/10/23 -- 10/27 & CIR\\ 
$\xi$ Per     &  9.4 & 52410 -- 52652 &  70 & 1994/10/15 -- 10/24 & CIR\\ 
HD 34656      &  5.1 & 40728 -- 40834 &  29 & 1991/01/03 -- 02/06 & weak 
\niv, $v=0$\\  
$\zeta$ Pup   &  5.3 & 36078 -- 36168 &  31 & 1989/04/24 -- 04/30 & large 
gaps, $v=0$ 
\\
$\zeta$ Pup   & 15.9 & 53338 -- 53783 & 149 & 1995/01/13 -- 01/29 & $v=0$, 
CIR?\\
HD 93843      & 28.1 & 57050 -- 57329 &  68 & 1996/05/05 -- 06/02 & poor 
sampling, not used \\ 
68 Cyg        &  6.6 & 28969 -- 29084 &  33 & 1986/08/23 -- 08/29 & large 
gaps, CIR?\\
68 Cyg        &  4.5 & 42787 -- 42920 &  40 & 1991/10/22 -- 10/27 & weak 
\niv, CIR? \\
68 Cyg        &  8.6 & 52420 -- 52650 &  66 & 1994/10/16 -- 10/24 & CIR\\
$\lambda$ Cep &  5.1 & 40727 -- 40833 &  24 & 1991/01/31 -- 02/05 & large 
gaps, $v=0$\\  
$\lambda$ Cep &  4.5 & 42786 -- 42919 &  40 & 1991/10/22 -- 10/27 & $v=0$, 
CIR?\\  
$\lambda$ Cep &  7.9 & 52425 -- 52635 &  32 & 1994/10/16 -- 10/24 & poor 
sampling, not used \\ \hline 
\end{tabular} 
\end{center}
\end{table}
\clearpage




\begin{thebibliography}{}

\bibitem[Cohen et al.\ (2014)]{co04} Cohen, D.H., Wollman, E.E., 
Leutenegger, M.A., Sundqvist, J.O., Fullerton, A.W., Zsarg\'{o}, J. \& 
Owocki, S.P.  2014, MNRAS, 439, 908

\bibitem[Cranmer \& Owocki (1996)]{co96} Cranmer, S.R. \& Owocki, S.P. 
1996, ApJ, 462, 469

\bibitem[de Jong et al.\ (2001)]{dj01} de Jong, J. A., Henrichs, H.F., 
Kaper, L., Nichols, J.S., Bjorkman, K., Bohlender, D.A., Cao, H., Gordon, 
K., Hill, G., Jiang, Y., Kolka, I., Morrison, N., Neff, J., O'Neal, D., 
Scheers, B. \& Telting, J.H. 2001, A\&A, 368, 601

\bibitem[Fullerton et al.\ (1997)]{f97} Fullerton, A.W., Massa, D.L., 
Prinja, R.K., Owocki, S.P. \& Cranmer, S.R. 1997, A\&A, 327, 699

\bibitem[Fullerton et al.\ (2006)]{f06} Fullerton, A.W., Massa, D.L. \& 
Prinja, R.K. 2006, ApJ, 637, 1025

\bibitem[Howarth et al.\ (1995)]{hpm95} Howarth, I.D., Prinja, R.K., Massa, 
D. 1995, ApJ, 452L, 65

\bibitem[Howarth \& Smith (1995)]{hs95}	Howarth, I.D. \& Smith, K.C. 1995, 
ApJ, 439, 431

\bibitem[Howarth \& Stevens (2014)]{hs14} Howarth, Ian D. \& Stevens, I-R. 
2014, MNRAS, 445, 2878 

\bibitem[Howarth \& Smith (1995)]{hs95} Howarth, I.D. \& Smith, K.C. 1995, 
ApJ, 439, 431

\bibitem[Kaper et al.\ (1996)]{k96} Kaper, L., Henrichs, H.F., Nichols, 
J.S., Snoek, L.C., Volten, H. \& Zwarthoed, G.A.A. 1996 A\&AS, 116, 257 
(k96)

\bibitem[Kaper et al.\ (1997)]{k97} Kaper, L., Henrichs, H.F., Fullerton, 
A.W., Ando, H., Bjorkman, K.S., Gies, D.R., Hirata, R., Kambe, E., McDavid, 
D. \& Nichols, J.S. 1997, A\&A, 327, 281 (k97)

\bibitem[Kaper et al.\ (1999)]{k99} Kaper, L., Henrichs, H.F., Nichols,J.S. 
\& Telting, J.H. 1999, A\&A, 344, 231 (k99)

\bibitem[Lobel \& Blomme (2008)]{lb08} Lobel, A. \& Blomme, R. 2008, ApJ, 
678. 408

\bibitem[Ma{\'i}z-Apell{\'a}niz et al.\ (2004)]{m04} Ma{\'i}z-Apell{\'a}niz, 
J., Walborn, N.R., Galu{\'e}, H.A., \& Wei, L.H. 2004, ApJS, 151, 103.

\bibitem[Massa et al.\ (1995)]{mega} Massa, D., Fullerton, A.W., Nichols, 
J.S., Owocki, S.P., Prinja, R.K., St-Louis, N., Willis, A.J., Altner, B., 
Bolton, C.T., Cassinelli, J.P. et al.\ 1995, ApJ, 452L, 53

\bibitem[Massa et al.\ (2000)]{m00} Massa, D., Fullerton, A.W., Hutchings, 
J.B., Morton, D.C., Sonneborn, G., Willis, A.J., Bianchi, L., Brownsberger, 
K.R., Crowther, P.A., Snow, T.P. \& York, D.G. 2000 ApJ, 538L, 47 

\bibitem[Massa et al.\ (2008)]{m08} Massa, D.L., Prinja, R.K., \& 
Fullerton, A.W. 2008, in Clumping in Hot-Star Winds, ed. W.-R. Hamann, A. 
Feldmeier, \& L. M. Oskinova, 147

\bibitem[Massa et al.\ (2014)]{m14} Massa, D., Oskinova, L., Fullerton, 
A.W., Prinja, R.K., Bohlender, D.A., Morrison, N.D., Blake, M. \& Pych, W. 
2014, MNRAS, 441, 2173

\bibitem[Muijres et al.\ (2013)]{mu13} Muijres, L.E., Vink, J.S., de Koter, 
A., M\"{u}ller, P.E. \& Langer, N. 2012, A\&A, 537A, 37

\bibitem[Mullan (1984)]{m84} Mullan, D.J. 1984, ApJ, 283, 303 

\bibitem[Naz\'{e} et al.\ (2013)]{n0384} Naz\'{e}, Y., Oskinova, L. M. \& 
Gosset, E. 2013, ApJ, 763, 143

\bibitem[Olson (1982)]{o82} Olson, G.L. 1982, ApJ, 255, 267 

\bibitem[Olson (1981)]{o81} Olson, G.L. 1981, ApJ, 245, 1054 

\bibitem[Owocki et al.\ (1995)]{oetal95} Owocki, S.P., Cramner, S.R. \& 
Fullert, A.W 1995, ApJ, 453, L37

\bibitem[Penny (1996)]{pen96} Penny, L.F. 1996, ApJ, 463, 737

\bibitem[Prinja (1988)]{prinja88} Prinja, R.K. 1988, MNRAS, 231P, 21

\bibitem[Prinja et al.\ (1992)]{petal92} Prinja, R. K., Balona, L.A., 
Bolton, C.T., Crowe, R.A., Fieldus, M.S., Fullerton, A.W., Gies, D.R., 
Howarth, I.D., McDavid, D. \& Reid, A.H.N. 1992, ApJ, 390, 266

\bibitem[Prinja \& Howarth (1988)]{ph88} Prinja, R.K., Howarth, I.D. 1988, 
MNRAS, 233, 123

\bibitem[Prinja et al.\ (1998)]{petal98} Prinja, R.K., Massa, D., Howarth, 
I.D. \& Fullerton, A.W. 1998, MNRAS, 301, 926

\bibitem[Prinja et al.\ (2002)]{petal02} Prinja, R.K., Massa, D. \& 
Fullerton, A.W. 2002, A\&A, 388, 587

\bibitem[Prinja \& Massa (2010)]{pm10} Prinja, R.K. \& Massa 2010, A\&A, 
521L, 55  

\bibitem[Prinja \& Massa (2013)]{pm13} Prinja, R.K. \& Massa 2013, A\&A, 
559A, 15

\bibitem[Prinja, Massa, \& Cantiello (2012)]{pmc13} Prinja, R.K., Massa, 
D.L. \& Cantiello, M. 2012, ApJ, 759L, 28

\bibitem[Ramiaramanantsoa et al.\ (2014)]{ram14} Ramiaramanantsoa, T., 
Moffat, A.F.J., Chen\'{e}, A-N, Richardson, N.D., Henrichs, H.F., 
Desforges, S., Antoci, V., Rowe, J.F., Matthews, J.M., Kuschnig, R., et 
al.\  2014, MNRAS, 441, 910

\bibitem[Sundqvist et al.\ (2010)]{sun10} Sundqvist, J.O., Puls, J. \& 
Feldmeier, A. 2010, A\&A, 510A, 11

\bibitem[S\v{u}rlan et al.\ (2013)]{sur13} S\v{u}rlan, B., Hamann, W.-R., 
Aret, A., Kub\'{a}t, J., Oskinova, L.M. \& Torres, A.F. 2013, A\&A, 559A, 
130 

\end{thebibliography}
\end{document}